\documentclass[reqno, 11pt]{amsart}

\usepackage{hyperref}

\usepackage{enumerate}

\usepackage{cite}

\numberwithin{equation}{section}

\usepackage[pdftex]{graphicx}

\newcommand\R{{\mathbb R}}

\newcommand\Z{{\mathbb Z}}
\newcommand\ve{\varepsilon}

\newcommand\cE{\mathcal E}

\usepackage{amssymb, amsmath}

\title[Energy Diffusion]{%
Energy Diffusion and Superdiffusion in Oscillators Lattice Networks}

\author{Stefano Olla}
\address{\!\!\!\!\!\!\!Ceremade, UMR CNRS 7534 \newline
  Universit\'e de Paris Dauphine,\newline
  Place du Mar\'echal De Lattre De Tassigny\newline
  75775 Paris Cedex 16 - France.\newline
\rm {\texttt{olla@ceremade.dauphine.fr}}\newline
 \texttt{http://www.ceremade.dauphine.fr/\~\;\!\!olla}
}

\date{\today}

\thanks{\textsc{Acknowledgements.}
The author acknowledge the support of the
 ANR LHMSHE n.BLAN07-2184264 and of the accord GREFI-MEFI}


\begin{document}

\begin{abstract}
  I review here some recent result on the thermal conductivity of
  chains of oscillators whose hamiltonian dynamics is perturbed by a
  noise conserving energy and momentum.
\end{abstract}

\maketitle

\section{Introduction}
\label{sec:introduction}

Let us consider a 1-dimensional chain of oscillators indexed by $x\in\Z$,
whose formal Hamiltonian is given by
\begin{equation}
  \label{eq:1}
  \mathcal H (p,q) = \sum_x \left[\frac{p_x^2}2 \ +\ V(q_{x+1} - q_x)
    + W(q_x)\right],
\end{equation}
where $q_x$ indicate the displacement of the atom $x$ from its
equilibrium position, and $p_x$ its momentum (we fix for the moment
all masses equal to 1). The potential $V$ and $W$ are some smooth positive
function growing at infinity fast enough. The $W$ potential is often
called \emph{pinning}.

We want to understand the \emph{macroscopic} properties of energy transport
for the corresponding hamiltonian dynamics
\begin{equation}
  \label{eq:2}
  \dot q_x = p_x, \qquad \dot p_x = - \partial_{q_x} \mathcal H. 
\end{equation}
When we say ``macroscopic energy transport'' we are meaning a certain
non-equilibrium evolution that we want to observe in a space--time
macroscopic scale that should be specified.  
This space--time scale can be typical of the model and the same model
can have distinct macroscopic scalings under which it will behave very
differently. For example in the case that $W=0$ (unpinned system),
total momentum is also conserved and hyperbolic scaling (in which
space and time are scaled in the same way) is a natural one. In the
hyperbolic scaling energy is carried around by the momentum of the
particles and macroscopic evolution equation is given by the Euler
equations. 

Another important scaling is the diffusive scaling (time scaled as square
of space) where transport of energy happens by diffusion. The
macroscopic equation is usually given by heat equation. For example,
when the pinning potential $W$ is present, total momentum is not
conserved and nothing moves in the hyperbolic scale. Energy will
\emph{move} on the diffusive space-time scale, and macroscopically its
evolution should be governed by heat equation. More precisely,
defining the empirical distribution of the energy
\begin{equation}
  \label{eq:3}
  \tilde\cE^\ve(G,t) = \ve \sum_x G(\ve x) \cE_x(\ve^{-2} t)
\end{equation}
where $\cE_x = \frac{p_x^2}2 \ +\ V(q_{x+1} - q_x) + W(q_x)$ is the
energy of particle $x$, and $G$ is a smooth test function on $\R$ with
compact support. We expect that
\begin{equation}
  \label{eq:4}
  \lim_{\ve\to 0} \tilde\cE^\ve(G,t) = \int G(y) u(y,t) \; dy
\end{equation}
in \emph{some statistical sense},
with $u(y,t)$ solution of the (non--linear) heat equation
\begin{equation}
  \label{eq:5}
  \partial_t u = \partial_{x} \left( \kappa (u)\partial_x u \right) \ .
\end{equation}
The function $\kappa (u)$ is called thermal conductivity and can be
expressed in terms of the dynamics in equilibrium:
\begin{equation}
  \label{eq:6}
 \kappa (u) = \lim_{t\to\infty} \frac 1{2t T^2} \sum_x x^2
  \left[\left< \cE_x(t) 
      \cE(0)\right>_T - u^2\right] 
\end{equation}
where the temperature $T=T(u)$
 correspond thermodynamically to the average energy $u$,
 $<\cdot>_T$
 is the expectation of the dynamics in equilibrium at temperature $T$.

Proving (\ref{eq:4})--(\ref{eq:5}) from an hamiltonian microscopic
dynamics is one of the major challenge in
non--equilibrium statistical mechanics \cite{BONETTO:2000p348}.
It is not clear under which conditions on the interaction and
initial conditions this result could be valid. Even
the proof of the existence of the limit (\ref{eq:6}) defining the
thermal conductivity is completely open for any deterministic system. 
 What is clear is that (\ref{eq:4})--(\ref{eq:5}) are not always
 valid. For example if $V$ and $W$ are quadratic (harmonic chains),
 then the energy corresponding to each Fourier mode is conserved and
 carried ballistically without any interaction with the other
 modes. It results that thermal conductivity is infinite in this case
 (\cite{Rieder:1967p389}, and for a macroscopic equation in
 the hyperbolic scaling see \cite{MR845729}, as explained in section
 \ref{sec:linet-limits:-phon}).   

In nonlinear unpinned cases ($W=0$) one expects, generically in
dimension 1,  that $\kappa = +\infty$, and correspondingly a
superdiffusion of energy. This fact seems confirmed by all numerical
simulation of molecular dynamics \cite{Lepri:1997p378}, even though
there is not a general agreement about the order of this
superdiffusivity. Most interesting would be to understand what kind of
 stochastic process would govern this superdiffusion.

Adding a stochastic perturbation to the hamiltonian dynamics certainly
helps to obtain some mathematical result in these problems. 
Of course adding noise to the microscopic
dynamics may change the macroscopic behavior. The ideal is to add
noise terms that change as little as possible the macroscopic
behaviour, at least qualitatively. For example it is important that
this noise conserves energy, and eventually momentum, since these are
the quantities we expect being \emph{conserved} in the infinite system.

\section{Conservative Stochastic Dynamics}
\label{sec:cons-stoch-dynam}

We consider the Hamiltonian dynamics weakly perturbed by a stochastic
noise acting only on momenta and locally preserving  momentum and
kinetic energy.
The generator of the dynamics is
\begin{equation}
L = A + \gamma S
\end{equation}
with $\gamma > 0$, where $A$ is the usual Hamiltonian vector field
\begin{equation}
\label{def:Agen}
\begin{split}
A = \sum_{yx\in\Z} \left\{ p_{x} \partial_{q_{x}} -
 (\partial_{q_x} \mathcal H) \; \partial_{p_{x}}\right\},
\end{split}
\end{equation}
while $S$ is the generator of the stochastic perturbation. The
operator $S$ acts only on the momenta $\{p_y\}$ and generates
a diffusion on the surface of constant kinetic energy and constant
momentum. $S$ is defined as
\begin{equation}\label{defq:Sgen1}
S = \frac 16 \sum_{z\in\Z}(Y_z)^2,
\end{equation}
where
\begin{equation*}
\label{eq:005}
Y_z=(p_z-p_{z+1})\partial_{p_{z-1}}+(p_{z+1}-p_{z-1})\partial_{p_z} +
(p_{z-1}-p_z)\partial_{p_{z+1}}
\end{equation*}
which is a vector field tangent to the surface of constant kinetic
energy and of constant momentum for three neighbouring particles. As
a consequence energy and momentum are locally conserved which, of course, implies also the conservation of total momentum and total
energy of the system, i.e. formally
\begin{equation*}
S \; \sum_{x\in\Z} p_x = 0\ ,\hspace{0.4cm} S \mathcal H = 0 .
\end{equation*}
Since also the hamiltonian dynamics conserves energy, we also have
$L\mathcal H = 0$. Furthermore in the unpinned case ($W = 0$), $L \;
\sum_{x} p_x = 0$.


The evolution of $\{p(t),q(t)\}$ is given by the following
stochastic differential equations
\begin{equation}
\label{eq:sde1}
\begin{split}
dq_x= & \;p_x \;dt ,\\
dp_x = & - \partial_{x} \mathcal H dt + \frac{\gamma} 6
\Delta(4p_x + p_{x-1} + p_{x+1}) dt \\ &\qquad\qquad
+ \sqrt{\frac{\gamma}3}
\sum_{k=-1,0, 1} \left(Y_{x+k} p_x \right) dw_{x+k}(t).
\end{split}
\end{equation}
Here $\{w_{y}(t)\}_{y\in\Z}$ are independent standard Wiener
processes and $\Delta$ is the discrete laplacian on $\Z$:
$\Delta f(z) = f(z+1) + f(z-1) -2f(z)$ .

In the unpinned 1-dimensional case, the equilibrium measures are
particularly simple. In fact the right coordinates are $r_x =
q_{x+1} - q_x$, and the family of product measures
\begin{equation}
  \label{eq:9}
  \mu_{\lambda, \pi, \beta} (dp,dr) = \prod_{x \in\Z}
  \frac{e^{-\beta(p_x-\pi)^2/2}}{\sqrt 2\pi}\ 
  \frac{e^{-\beta V(r_x) + \lambda r_x}}{Z(\lambda, \beta)} 
\end{equation}
are stationary for the dynamics. The three parameters $\lambda, \pi,
\beta$ correspond to the 3 conserved quantities of the dynamics
(energy, momentum and $\sum_x r_x$, the \emph{stretch} of the
chain), while $Z(\lambda,\beta)$  is the normalization constant. 
It can be proven that these are the only translation invariant
stationary measures of the dynamics (\cite{Olla:2008notes}, we call
this property ``ergodicity of the infinite dynamics''). In more
dimensions the unpinned case is much more complex and even the
definition of these equilibrium measures are problematic
(cf. \cite{Funaki:1997p1830}).

In the unpinned one-dimensional case, since momentum is conserved,
there is a non trivial macroscopic evolution on the conserved
quantities in the hyperbolic scaling (space and time scaled in the
same way). Let us define the energy of particle $x$ as
\begin{equation*}
 \cE_x = \frac{p_x^2}2 + \frac 12 \left(V(r_{x-1}) + V(r_x)\right)
\end{equation*}
Locally the conservation of energy can be written as
\begin{equation}
  \label{eq:7}
  L  \cE_x = j^{\cE}_{x-1,x} -j^{\cE}_{x,x+1}
\end{equation}
where the instantanueous current of energy $j^{\cE}_{x,x+1}$ is the sum
of the current due to the hamiltonian mechanism plus the current due
to the stochastic term of the dynamics:
\begin{equation}
  \label{eq:8}
  \begin{split}
    j^{\cE}_{x,x+1} = j^{\cE,a}_{x,x+1} & + j^{\cE,s}_{x,x+1} \\
     j^{\cE,a}_{x,x+1} = - \frac 12 \left(p_x + p_{x+1}\right)
     V'(r_x),& \qquad
      j^{\cE,s}_{x,x+1} = -\gamma \nabla \varphi_x
  \end{split}
\end{equation}
with $\varphi_x= (p_{x+1}^2 + 4p_{x}^2 + p_{x-1}^2 + p_{x+1}p_{x-1} -
2p_{x+1}p_{x} - 2 p_{x}p_{x-1})$. Similarly conservation of momentum
reads as
\begin{equation}
  \label{eq:10}
  \begin{split}
     L  p_x &= j^{p}_{x-1,x} -j^{p}_{x,x+1}, \\
     j^p_{x,x+1} &= V'(r_x) + \frac{\gamma}{6} \nabla (4p_x + p_{x-1} + p_x) 
  \end{split}
\end{equation}
while mass conservation is simply given by
\begin{equation}
  \label{eq:11}
  L r_x = p_{x+1} - p_x .
\end{equation}

Let $G(y)$ a test function continuous with compact support. We expect
that 
\begin{equation}\label{eq:hhd}
  \begin{split}
     \epsilon \sum_x G(\epsilon x)
    \begin{pmatrix}
      r_x(\epsilon^{-1} t) \\ p_x(\epsilon^{-1} t)\\ \cE_x(\epsilon^{-1} t)
    \end{pmatrix}
    \ \ \mathop{\longrightarrow}_{\epsilon\to 0}^{\text{probability}}\ \ \int G(y)  
    \begin{pmatrix}
      \mathfrak r(t,y) \\ \mathfrak p(t,y)\\ \mathfrak e(t,y)
    \end{pmatrix}
    dy
  \end{split}
\end{equation}
where $\mathfrak r(t,y), \mathfrak p(t,y), \mathfrak e(t,y)$ are given
by the solution of the Euler hyperbolic system of equations
\begin{equation}
  \label{eq:euler}
   \begin{split}
    \partial_t \frak r &= \partial_y \frak p\\
    \partial_t \frak p &= \partial_y P(\frak r, \frak e - \frak p^2/2)
    \\
    \partial_t \frak e &=  \partial_y\left(\frak p P(\frak r, \frak e -
      \frak p^2/2) \right) 
  \end{split}
\end{equation}
Here $P(r,u)$ is the thermodynamic pressure, which is related to the
thermodynamic entropy $S(r,u)$ by the relation
\begin{equation}
  \label{eq:12}
  P(r,u) = -\frac{\partial_r S(r,u)}{\partial_u S(r,u)}
\end{equation}
and $S$ is defined from $Z(\lambda, \beta)$ with a Legendre transform:
\begin{equation}
  \label{eq:13}
  S(r,u) = \sup_{\lambda, \beta} \left\{ \lambda r -\beta u -
  \log\left( Z(\lambda, \beta) \sqrt{\beta/2\pi}\right) \right\}
\end{equation}
Pressure $P(r,u)$ is also give by the expectation of $V'(r_x)$ with
respect to $\mu_{\lambda,\pi,\beta}$, for the corresponding values of
the parameters $\lambda$ and $\beta$. 

The hydrodynamic limit (\ref{eq:hhd}) can be proven rigorously in the
smooth regime of equation (\ref{eq:euler}), by using the relative entropy
method (\cite{Olla:1993p69}, see also \cite{Basile:2007Phd} and
\cite{Olla:2008notes} for the application to this specific model).
Note that (\ref{eq:hhd}) does not depend on the strength of the
microscopic noise $\gamma$. Noise here is used only to prove some
ergodic properties for the dynamics, necessary to obtain the
result. In fact without noise this limit may not be true, as for
example in the linear case ($V$ quadratic, see below). 
Note also that for smooth solutions the macroscopic evolution
(\ref{eq:euler}) is locally isoentropic, i.e.
\begin{equation}
  \label{eq:14}
  \frac d{dt} S\left(\mathfrak r(t,y), \mathfrak e(t,y) - \frac{\mathfrak
  p(t,y)^2}2 \right) = 0
\end{equation}
A challenging open problem is to extend this result to solutions that
present shocks, where the above derivative is (presumably) strictly positive.


\section{Diffusive evolution: Green-Kubo Formula}
\label{sec:green-kubo-formula}

In the pinned model ($W>0$) momentum is not conserved, energy is the
only relevant conserved quantity for the infinite system and the
equilibrium measure are the Gibbs measures at given temperature,
corresponding to the hamiltonian $\mathcal H$. These probability
measures are defined by the usual DLR
equations. Consequently the hyperbolic scaling is trivial (nothing
moves at that time scale). In order to see energy moving at a
macroscopic scale, one has to look at larger time scale. The natural
scaling is the diffusive one: we expect, for a given test function $G$
as above, 
\begin{equation}
  \label{eq:15}
   \epsilon \sum_x G(\epsilon x) \cE_x(\epsilon^{-2} t) 
    \mathop{\longrightarrow}_{\epsilon\to 0}^{\text{probability}}\ \ \int G(y)  
      \mathfrak T(t,y)\; dy
\end{equation}
where $ \mathfrak T(t,y)$ is the solution of the (non-linear) heat equation
\begin{equation}
  \label{eq:16}
  \partial_t \mathfrak T = \partial_{x}\left(\kappa(\mathfrak
    T)\partial_x \mathfrak T \right), 
\end{equation}
where $\kappa(T)$ is the thermal conductivity at temperature $T$. This is
given by the Green-Kubo formula  
\begin{equation}
  \label{eq:17}
  \kappa(T) =\frac 1{2\chi(T)} \left[\int_0^\infty
    \sum_{x=-\infty}^{+\infty}  \left<  j^{\cE,a}_{x,x+1} (t)
    j^{\cE,a}_{0,1}(0) \right>_T \; dt + \gamma T^2\right] 
\end{equation}
where $ \left<  j^{\cE,a}_{x,x+1} (t)  j^{\cE,a}_{0,1}(0) \right>_T$
denote the expectation with respect to the dynamics in equilibrium at
temperature $T = \beta^{-1}$.  
The explicit $\gamma T^2/2\chi(T)$ term is the contribution of the
stochastic part of the dynamics. Formula (\ref{eq:17}) can be obtained
from (\ref{eq:6}) using the conservation of energy (see
\cite{Basile:2006p134} for a proof).
  We believe that such statement is
always true for non--linear pinned dynamics, also in the deterministic case
($\gamma = 0$). But
even for $\gamma >0$, this is hard to prove and
still an open problem. Even the convergence of the integrals defining  
(\ref{eq:17}) is not known if non-linearities are present (some bounds
are proven in \cite{Basile:2006p134}). The only case with non-linear
interaction in which
(\ref{eq:17}) is proven convergent is when the noise is generated by
Langevin heat bath attached at each particle of the system
\cite{Bonetto:2008}, a non conservative stochastic perturbation. 

In unpinned systems the situation is more complex because of the
momentum conservation. To avoid complicate re-centering along
characteristics of (\ref{eq:euler}), we can consider initial
random configurations with momentum of (locally) zero average and
constant density profile, only gradients of temperature admitted. Then
conductivity $\kappa$ is also a function of the density of particles and in
formula (\ref{eq:17}) the expectation should be taken with respect to
the equilibrium dynamics with $\pi = 0$ (i.e. starting with
configurations distributed by $\mu_{\lambda, 0, \beta}$ defined by
  (\ref{eq:9})). Numerical evidence shows that in this one-dimensional
  case $\kappa = +\infty$ (\cite{Lepri:1997p378}, also for $\gamma
  >0$, as long as 
  momentum is conserved \cite{Basile:2007epstx}). In the physics
  literature there is 
  a long discussion about the nature and the \emph{order} or this
  superdiffusion. From dimension 3 on, it is expected that formulas
  corresponding to (\ref{eq:17}) give a finite diffusivity.

Rigorous results can be proven for the harmonic case with $\gamma>0$
\cite{Basile:2006p1127, Basile:2006p134}. It turns out that  $\kappa$
 is finite if system is pinned or in dimension $d\ge 3$, while the 1
and 2 dimensional unpinned cases are superdiffusive.

\section{Kinetic Limits: Phonon Boltzmann Equation}
\label{sec:linet-limits:-phon}

The harmonic case is enough simple to obtain some non-equilibrium
results. 
In \cite{Basile:2008p1473} we consider the hyperbolic scaling in a
weak noise limit. Noise is rescaled by multiplying its strength
$\gamma$ by $\epsilon$. This way the effect of the noise per particle
remains finite in the macroscopic scale (
that motivates
the term \emph{kinetic} in defining this limit). This  is in the same
spirit as the model of hard sphere with random collision considered in
\cite{MR2076921}. The right quantity to 
look here is the Wigner distribution of the energy, formally defined
as
\begin{equation}
  \label{eq:18}
  W^{\epsilon} (y,k,t) = \frac{1}2
  \int_{-1/2}^{1/2} e^{i 2\pi y\eta/\epsilon}
  \left<\psi(k-\eta/2, t/\epsilon)^* \psi(k+\eta/2,
    t/\epsilon) \right> \; d\eta
\end{equation}
where
\begin{equation*}
  \psi(k,t) = \frac 1{\sqrt 2} \left(\omega(k) \hat q (k,t) + i \hat
    p(k,t) \right),
\end{equation*}
here $\hat q(k,t), \hat p(k,t)$ are the Fourier transform of the $q_y(t),
p_y(t)$, and $\omega(k)$ is the dispersion relation of the lattice, which
in this one-dimensional nearest neighbour case is given by
$\omega (k) = c|\sin (\pi k)|$ (\emph{acoustic dispersion}).
The result in  \cite{Basile:2008p1473} states that $W^{\epsilon}
(y,k,t)$ converges, as a distribution on $\mathbb R\times [0,1]$, to the
solution of the linear transport equation
\begin{equation}
  \label{eq:19}
  \partial_t W(y,k,t)
  +\frac{\omega'(k)}{2\pi}\partial_y W(y,k,t)=
  \gamma\int C(k,k') \left(W(y,k',t) - W(y,k,t)\right) \; dk \,. 
\end{equation}
In the deterministic case ($\gamma = 0$) this result was obtained by
Dobrushin et al. in \cite{MR845729}, see also \cite{Mielke2006}.
The collision kernel $C(k,k')$ is positive and symmetric. It is
computable explicitly (cf.  \cite{Basile:2008p1473}), but the
important point is that $C(k,k') \sim k^2$ for small $k$. 
This is a consequence of the conservation of momentum: long waves
scatter very rarely.
Because $C(k,k')$ is positive, (\ref{eq:19}) has a simple
probabilistic interpretation: $W$ is the density at time $t$ of the energy
of particles (phonons) of mode $k$. 
A phonon of mode $k$ moves with
velocity $\omega'(k)$ and after an exponentially distributed random
time of intensity $\gamma C(k,k')$ changes its mode to $k'$.
Defining a Markov jump process $K(t)$ in $[0,1]$ with jumping rate
$\gamma C$, the position of the phonons is given by $Y(t) =
\int_0^t \omega'(K(s)) ds$. 

Thermal conductivity can be computed from (\ref{eq:19})
(cf. \cite{Basile:2008p1473}) and the results
are in accord with the direct calculations done in
\cite{Basile:2006p134}.  

\section{Levy's Superdiffusion of Energy}
\label{sec:levys-superd-energy}

As we mention in the previous section, phonons of small $k$ scatter
rarely, but their velocity $\omega'(k)$, in the unpinned case, are
still of order 1 as $k\to 0$. This induces a superdiffusive behavior of
these phonons. In    
\cite{Jara:2008p1191}, as application of new limit theorems for
functionals of Markov chains and processes, we prove that, for $\alpha
= 3/2$,
\begin{equation}
  \label{eq:20}
  \epsilon Y( \epsilon^{-\alpha} t) \
  \mathop{\longrightarrow}_{\text{law}}\ \mathcal L(t)
\end{equation}
where $\mathcal L(t)$ is a Levy $\alpha$-stable process, i.e. a
stochastic process with independent stationary increments and with
$\mathcal L(1)$ distributed by a $\alpha$-stable law. In terms of the
solution $W(y,k,t)$ of the equation (\ref{eq:19}) this result implies
the convergence
\begin{equation}
  \label{eq:21}
  \lim_{\epsilon\to 0} \int \left| W(\epsilon^{-\alpha} t, \epsilon^{-1}y,
    k) - \bar u(t,y)\right|^2 \; dk = 0
\end{equation}
where $\bar u(t,y)$ is the solution of the fractional heat equation
\begin{equation}
  \label{eq:22}
  \partial_t \bar u = -c ( -\Delta_y)^{\alpha/2} \bar u \ .
\end{equation}
where $c$ is a positive constant.

In the pinned case all the above result are still valid, but since the
velocity of the phonons
$\omega'(k) \sim k$ for small k, we have a regular diffusive behavior,
and $\alpha = 2$. 

It would be very interesting to understand how these results extends
to the anharmonic cases. Equation (\ref{eq:19}) will be substituted by
the non-linear phonon Boltzmann equation \cite{MR2264633}. In the
unpinned one-dimensional case thhis equation still will produce a
superdiffusion. Is it again of Levy type, or will have some
non-markovian terms?


 \bibliography{icmpbiblio}{}

\begin{thebibliography}{10}

\bibitem{Basile:2007Phd}
G~Basile.
\newblock PhD thesis, Universit\'e Paris Dauphine, 2007.

\bibitem{Basile:2007epstx}
G~Basile, L~Delfini, S~Lepri, R~Livi, S~Olla, and A~Politi.
\newblock Anomalous transport and relaxation in classical one-dimensional
  models.
\newblock {\em Eur. Phys. J. Special Topics}, 2007.

\bibitem{Basile:2006p1127}
Giada Basile, C{\'e}dric Bernardin, and Stefano Olla.
\newblock Momentum conserving model with anomalous thermal conductivity in low
  dimensional systems.
\newblock {\em Phys. Rev. Lett.}, 96(20):4, May 2006.

\bibitem{Basile:2006p134}
Giada Basile, Cedric Bernardin, and Stefano Olla.
\newblock Thermal conductivity for a momentum conserving model.
\newblock {\em arXiv}, cond-mat.stat-mech, Jan 2006.
\newblock New version with bounds for the anharmonic case.

\bibitem{Basile:2008p1473}
Giada Basile, Stefano Olla, and Herbert Spohn.
\newblock Wigner functions and stochastically perturbed lattice dynamics.
\newblock {\em arXiv}, math.PR, May 2008.

\bibitem{Bonetto:2008}
F~Bonetto, J~Lebowitz, J~Lukkarinen, and S~Olla.
\newblock Heat conduction and entropy production in anharmonic crystals with
  self-consistent stochastic reservoirs.
\newblock {\em arXiv}, 2008.

\bibitem{BONETTO:2000p348}
F~Bonetto, J~Lebowitz, and L~Rey-Bellet.
\newblock Fourier's law: A challenge to theorists.
\newblock {\em Mathematical Physics 2000}, Jan 2000.

\bibitem{MR845729}
R.~L. Dobrushin, A.~Pellegrinotti, Yu.~M. Suhov, and L.~Triolo.
\newblock One-dimensional harmonic lattice caricature of hydrodynamics.
\newblock {\em J. Statist. Phys.}, 43(3-4):571--607, 1986.

\bibitem{Funaki:1997p1830}
T~Funaki and H~Spohn.
\newblock Motion by mean curvature from the ginzburg-landau interface model.
\newblock {\em Comm. Math. Phys.}, Jan 1997.

\bibitem{Jara:2008p1191}
Milton Jara, Tomasz Komorowski, and Stefano Olla.
\newblock Limit theorems for additive functionals of a markov chain.
\newblock {\em arXiv}, math.PR, Sep 2008.

\bibitem{Lepri:1997p378}
S~Lepri, R~Livi, and A~Politi.
\newblock Heat conduction in chains of nonlinear oscillators.
\newblock {\em Phys. Rev. Lett.}, Jan 1997.

\bibitem{Mielke2006}
Alexandre Mielke.
\newblock Macroscopic behavior of microscopic oscillations in harmonic lattices
  via wigner-husimi transforms.
\newblock {\em Arch. Rat. Mech. Anal.}, 181:401--448, 2006.

\bibitem{Olla:2008notes}
S~Olla.
\newblock Notes cours ihp.

\bibitem{Olla:1993p69}
S~Olla, S.~R.~S Varadhan, and H.-T Yau.
\newblock Hydrodynamical limit for a hamiltonian system with weak noise.
\newblock {\em Comm. Math. Phys.}, 155(3):523--560, 1993.

\bibitem{MR2076921}
Fraydoun Rezakhanlou.
\newblock Boltzmann-{G}rad limits for stochastic hard sphere models.
\newblock {\em Comm. Math. Phys.}, 248(3):553--637, 2004.

\bibitem{Rieder:1967p389}
Z~Rieder, J.~L Lebowitz, and E~Lieb.
\newblock Properties of a harmonic crystal in a stationary nonequilibrium
  state.
\newblock {\em Journal of Mathematical Physics}, 8:1073, May 1967.

\bibitem{MR2264633}
Herbert Spohn.
\newblock The phonon {B}oltzmann equation, properties and link to weakly
  anharmonic lattice dynamics.
\newblock {\em J. Stat. Phys.}, 124(2-4):1041--1104, 2006.

\end{thebibliography}
 \bibliographystyle{plain}

\end{document}